\documentclass[twocolumn,showpacs,preprintnumbers,prd,floatfix]{revtex4}
\usepackage{graphicx}

\begin{document}
\preprint{ITEP-TH-37/07}

\title{Renormalization of the vacuum angle in quantum mechanics, \\
Berry phase and continuous measurements}

\author{S. M. Apenko }

\affiliation{ Theory Department, Lebedev Physics Institute,
Moscow, 119991, Russia}
 \altaffiliation[Also at ]{ ITEP, Moscow, 117924, Russia}
 \email{apenko@lpi.ru}

\date{\today}

\begin{abstract}
The vacuum angle $\theta$ renormalization is studied for a toy
model of a quantum particle moving around a ring, threaded by a
magnetic flux $\theta$. Different renormalization group (RG)
procedures lead to the same generic RG flow diagram, similar to
that of the quantum Hall effect. We argue that the renormalized
value of the vacuum angle may be observed if the particle's
position is measured with finite accuracy or coupled to additional
slow variable, which can be viewed as a coordinate of a second
(heavy) particle on the ring. In this case the renormalized
$\theta$ appears as a magnetic flux this heavy particle sees, or
the Berry phase, associated with its slow rotation.
\end{abstract}

 \maketitle

In quantum field theories  it is sometimes possible (e.g. in
quantum chromodynamics (QCD) or non-linear $\sigma$ models) to add
the so-called topological term to the action and to consider the
coefficient $\theta$ in front of this term, usually called vacuum
or topological angle, as an additional parameter of the theory
(see, e.g. \cite{Pol}). Long ago it was suggested, that vacuum
angle $\theta$  becomes scale dependent (as any other running
coupling constant) if properly defined (non-perturbative)
renormalization group (RG) transformation is introduced
\cite{KM,P}, and flows to zero (mod $2\pi$) in the infrared limit
(see also \cite{rec} for some recent works). Hence one might
expect that the observable low energy $\theta$ should vanish,
possibly solving the strong CP problem in QCD (i.e. why we do not
observe CP violation due to the $\theta$-term while {\em a priori}
there are no reasons to put $\theta=0$ \cite{Sh}). But such a
renormalization is, in a sense, counter-intuitive, since $\theta$
more resembles some quantum number (and is related to a
superselection rule) than usual coupling constant. Moreover,
non-perturbative calculations, based on the sum rule approach
\cite{Sum} (see also \cite{Sh} and references therein) have shown
that CP violating effects actually depend on the bare $\theta$, so
that it is not clear what does the $\theta$ renormalization
actually mean in QCD and how it may be observed.

Perhaps the most known example where such renormalization have
proved to be important is the quantum Hall effect (QHE). In this
case, described by a matrix non-linear $\sigma$ model, the
renormalized vacuum angle is in fact defined as the observable
Hall conductivity, dependent on the sample's size or temperature
(see e.g. \cite{P,H}).

Quite recently it became clear that charging effects in a single
electron box (a metallic island coupled to the outside circuit by
a tunnel junction), also described by a topological term, are
closely related to the $\theta$ renormalization \cite{B,A}. This
last model is equivalent to ordinary quantum mechanics of a
particle (with friction in general case) on a ring  threaded by a
magnetic flux $\theta$, which can serve as the simplest zero
dimensional toy model to study the $\theta$ renormalization in
more detail.

It is possible to introduce a RG transformation in quantum
mechanics, similar in spirit to decimation procedure in one
dimensional classical spin models and related to continuous
measurements theory, which leads to the $\theta$ renormalization
of the required type \cite{A}, which manifest itself, as in QHE,
as temperature dependence of a certain observable. Renormalization
of $\theta$ is seen then to follow from the loss of information
about the initial topological charge in the course of the RG
transformation. The RG scheme of Ref. \cite{A} is, however,
somewhat artificial, since as a first step it introduces a lattice
(like time slices in the Trotter decomposition, used e.g. in path
integral Monte Carlo calculations \cite{Cep}) to be removed in the
end.

For this reason here we present a different RG approach, also
inspired by an analogy between RG and continuous measurements, but
with no lattice and at zero temperature. Now the renormalized
$\theta$ appears as an effective magnetic flux  seen by an
additional slow variable (or Berry phase, related to its cyclic
evolution, compare with \cite{AW}). The resulting RG flow diagram
again has the typical QHE-like form with $\theta$ going to zero
(mod $2\pi$) in the infrared limit. Physical reasons for such
behaviour are also discussed.

Consider a particle of mass $m$ moving around a ring of unit
radius threaded by a magnetic flux $\theta$ (in units
$c=\hbar=e=1$). The corresponding (euclidian) action at finite
temperature may be written in terms of a planar unit vector $\bf
{n}(\tau)$ (${\bf n}^2=1$) which depends on a one-dimensional
coordinate (euclidean time)
\begin{equation}\label{A}
  S_0[{\bf n}]=\frac{m}{2}\int_{0}^{\beta}\dot{\bf{n}}^2(\tau)d\tau-
  i\frac{\theta}{2\pi}\int_{0}^{\beta}\epsilon_{ab}n_{a}(\tau)\dot{n}_{b}(\tau)d\tau,
\end{equation}
where $\epsilon_{ab}$ is the two dimensional antisymmetric tensor
and $\beta$ is the inverse temperature (we will assume
$\beta\rightarrow\infty$ in what follows). Since ${\bf n}(0)={\bf
n}(\beta)$ the model is actually defined on a circle. The last
term in (\ref{A}) has the form $i\theta Q$ where $Q$ is the
topological charge which distinguishes inequivalent mappings
$S^1\rightarrow S^1$ and takes integer values (equal to a number
of  rotations the particle make in time $\beta$), making the
theory periodic in $\theta$.

The magnetic flux $\theta$ explicitly breaks T invariance, the
most obvious T-violating effect being the non-zero persistent
current in the ground state. This is the analog of the CP problem
in QCD and now one may ask, how the dependence on $\theta$ can be
removed. One possible answer is that the magnetic flux could be
screened, if we allow the back reaction of the current on
$\theta$. This may be done by introducing an additional dynamical
variable (axion), coupled to the topological charge density.
Curiously, the model (\ref{A}) with the axion have been introduced
in a different context to describe a shunted Josephson junction
\nolinebreak \cite{J}.

Suppose now that we perform a continuous monitoring of the
particle position (in euclidean time) with a finite accuracy. If a
continuous quantum measurement results in a smooth slowly varying
trajectory ${\bf n}_0(\tau)$ then the corresponding amplitude may
be obtained through the restricted path integral \cite{M}
\begin{equation}\label{P}
  U[{\bf n}_0]=\int
  D{\bf n}(\tau)  \delta({\bf n}^{2}(\tau)-1)
  w[{\bf n},{\bf n}_0]\exp(-S_0[{\bf n}]),
\end{equation}
where the weight functional $w[{\bf n},{\bf n}_0]$ is usually
taken in a simple Gaussian form
\begin{equation}\label{G}
  w[{\bf n},{\bf n}_0]=\exp\left(-\frac{\lambda}{2}\int_0^{\beta}[{\bf n}(\tau)-{\bf
  n}_0(\tau)]^2 d\tau \right)
\end{equation}
and the constant $\lambda$ determines the accuracy of the
measurement.

Integration in Eq. (\ref{P}) defines an effective action
$U\sim\exp(-S_{eff}[{\bf n}_0])$ and hence a generalized Wilsonian
RG transformation with all coupling constants running with
$\lambda$. If we e.g. apply the same prescription to the 2D $O(N)$
$\sigma$ model then in the one-loop calculation of Ref. \cite{N}
$\lambda$ effectively acts as a mass squared for Goldstone modes,
leading thus to the charge renormalization
$\sim\ln(\Lambda/\sqrt{\lambda})$ ($\Lambda$ is the ultraviolet
cutoff). Hence changing $\lambda$ is indeed similar to changing
the scale. We now argue, that beyond the perturbation theory
$\lambda$ also may be viewed as a scale parameter.

For $\lambda$ large enough only paths close to ${\bf n}_0(\tau)$
contribute to the path integral (\ref{P}). But for the particle on
the ring it is possible that a given path ${\bf n}(\tau)$ is close
to ${\bf n}_0(\tau)$ for the most of the time, but suddenly makes
a fast complete rotation around the ring in time $\tau_0$. For
such instanton-like paths the weight factor (\ref{G}) behaves as
$w\sim\exp(-{\rm const}\times\lambda\tau_0)$, so that
``instantons'' with size $\tau_0>1/\lambda$ are strongly
suppressed (very fast rotations with $\tau_0\ll m$ are suppressed
by the kinetic term in Eq. (\ref{A})). Then with decreasing
$\lambda$ more and more instanton-like paths of larger scale
contribute to the integral (\ref{P}). Clearly, this is exactly
what a physicist usually expects from the RG transformation in
theories with instantons.

If we combine the action (\ref{A}) with the exponential from
(\ref{G}) then  the resulting action in (\ref{P}) (up to a
constant)
\begin{equation}\label{S}
S[{\bf n}]=S_0[{\bf n}]+\lambda\int_0^{\beta}{\bf n}(\tau){\bf
n}_0(\tau) d\tau
\end{equation}
describes the particle on the ring in time dependent electric
field $\lambda {\bf n}_0(\tau)$. For slowly varying ${\bf
n}_0(\tau)$ at zero temperature one can treat this problem in the
adiabatic approximation. Then, if the electric field makes one
complete revolution, the ground state will turn back to itself up
to a phase factor (Berry phase \cite{Ber}) which we denote by
$\exp(i\theta')$. If we introduce polar angles $\phi$ and $\phi_0$
instead of the vectors ${\bf n}$ and ${\bf n}_0$ then the
corresponding Hamiltonian may be written as
\begin{equation}\label{H}
  H=\frac{1}{2m}\left(-i\frac{{\partial}}{{\partial
  }\phi}-\frac{\theta}{2\pi}\right)^2+\lambda\cos(\phi-\phi_0(t))
\end{equation}
Let $\psi_0(\phi)=\psi_0(\phi-\phi_0)$ be the instantaneous ground
state wavefunction for the Hamiltonian (\ref{H}) with the energy
$E_0$, which obviously does not depend on $\phi_0$. Then the Berry
phase for the adiabatic change of $\phi_0$ from zero to $2\pi$ is
given by  \cite{Ber}
\begin{equation}\label{B}
  \theta'=i\int_0^{2\pi}d\phi_0
  \langle\psi_0|\frac{\partial}{\partial\phi_0}|\psi_0\rangle
\end{equation}
Since $\psi_0$ depends only on the difference $\phi-\phi_0$ we
have
\begin{eqnarray}\label{av}
&&\langle\psi_0|\frac{\partial}{\partial\phi_0}|\psi_0\rangle=
-\langle\psi_0|\frac{\partial}{\partial\phi}|\psi_0\rangle=\nonumber\\
&&-\langle\psi_0|\left(\frac{\partial}{\partial\phi}-i\frac{\theta}{2\pi}\right)|\psi_0\rangle
-i\frac{\theta}{2\pi}
\end{eqnarray}
The first term on the r.h.s. of Eq. (\ref{av}) is proportional to
the average of the derivative $\partial H/\partial\theta$ and
hence
\begin{equation}\label{b}
  \theta'=\theta-4\pi^2 m\frac{\partial E_0}{\partial\theta}
\end{equation}
The nontrivial Berry phase, different from $\theta$, means that
the coarse grained continuously measured trajectory sees a
``renormalized'' magnetic field, as was discussed in \cite{A}, due
to unobservable fast instanton-like rotations. This implies that
for slowly varying ${\bf n}_0$ we should have
\begin{eqnarray}\label{U}
 & U[{\bf n}_0]\sim &\exp\{-i\frac{\theta'}{2\pi}
 \int_{0}^{\beta}\epsilon_{ab}n_0^{a}(\tau)\dot{n}_0^{b}(\tau)d\tau+
 \nonumber\\
 &&+\frac{m'}{2}\int_{0}^{\beta}\dot{{\bf n}}_0^2(\tau)d\tau+\ldots\},
\end{eqnarray}
where dots indicate terms with higher derivatives of ${\bf n}_0$
and higher powers of $\dot{{\bf n}}_0$ and the renormalized mass
will be determined below.

Similar origin of topological terms from a corresponding Berry
phase was discussed in detail in Refs. \cite{AW} where fermions
were coupled to the background vector field in various space-time
dimensions (fermionic $\sigma$-models). Then integration over
fermions results in Eq. (\ref{U}) for planar vector ${\bf n}_0$
with $\theta'$, $m'$ dependent on the coupling constants. Here the
fast mode which is integrated out is also the planar vector, so
that it is more natural to speak of the $\theta$ renormalization
rather than of the induced topological term.

\begin{figure}[t]
\includegraphics[width=3in]{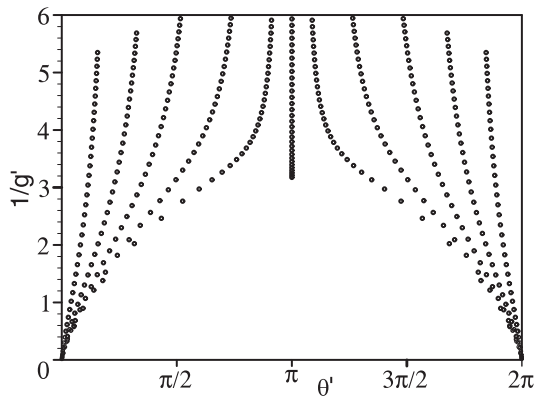}
\caption{Renormalized parameters $1/g'=\sqrt{m'\lambda}$ and
$\theta'$ from Eqs. (\ref{b}), (\ref{m}) for different values of
initial $\theta$. $\lambda$ decreases from top to bottom.}
\label{fig1}
\end{figure}

There exists a simple heuristic way to derive the expansion of Eq.
(\ref{U}). Consider a reference frame rotating with an angular
frequency $\omega=\dot{\phi}_0$, which is assumed to be small and
almost constant. In this frame ${\bf n}_0$ is constant, but an
additional magnetic field $2m\omega$ is present according to the
Larmor's theorem. Hence the hamiltonian $H'$ in the rotating frame
should be taken at the shifted value of the vacuum angle
$\theta+2m\pi\omega$, or more precisely,
\begin{equation}\label{er}
  H'=H+i\omega\frac{\partial}{\partial\phi}=
  H(\theta+2m\pi\omega)-\frac{\theta}{2\pi}\omega-\frac{m}{2}\omega^2
\end{equation}
(see e.g. \cite{ring}), where the last term is the centrifugal
potential (for the thin ring of unit radius) and the second one is
due to the presence of the magnetic flux $\theta$. Then if the
particle is in its ground state the effective action (after Wick
rotation $t\rightarrow -i\tau$ and expansion in powers of
$\dot{\phi}_0$ ) may be written as
\begin{eqnarray}\label{eff}
  &S_{eff}\simeq &\int_0^{\beta}d\tau \left[\frac{m}{2}\dot{\phi}_0^2-
  i\frac{\theta}{2\pi}\dot{\phi}_0+E_0(\theta+2mi\pi\dot{\phi}_0)\right]=
  \nonumber\\
  &&={\rm const}+\int_0^{\beta}d\tau \left[\frac{m'}{2}\dot{\phi}_0^2-
  i\frac{\theta'}{2\pi}\dot{\phi}_0+\ldots\right],
\end{eqnarray}
where $\theta'$ is given by the previously derived formula
(\ref{b}) and
\begin{equation}\label{m}
  m'=m-4m^2\pi^2\frac{\partial^2E_0}{\partial\theta^2}
\end{equation}
Clearly, this is the same action as in Eq. (\ref{U}). Formulas
(\ref{b}) and (\ref{m}) look very similar to the RG equations
derived in \cite{A}. Note, that they are independent of the
specific form of the coupling between ${\bf n}$ and ${\bf
n}_0$---all details are hidden in the ground state energy
$E_0(\theta)$.

For large $\lambda$, when the effective electric field is strong,
the $\theta$ dependence of $E_0$ is suppressed and
$\theta'\simeq\theta$. In this case $E_0$ depends on $\theta$ only
through instantons, as discussed in detail in \cite{R}, and
\begin{equation}\label{E}
  E_0(\theta)\simeq{\rm const}-2\sqrt{S_0}K{\rm e}^{-S_0}\cos\theta
\end{equation}
where $S_0(\lambda)\sim \sqrt{m\lambda}$ is the classical
instanton action and $K=K(\lambda)$ results from the ratio of
determinants \cite{R}. Then in terms of dimensionless ``coupling
constants'' $g=1/\sqrt{m\lambda}$ and $g'=1/\sqrt{m'\lambda}$ we
finally have at $g\rightarrow 0$
\begin{eqnarray}\label{ren}
  \theta'\simeq \theta -D(g){\rm
  e}^{-c/g}\sin\theta, \nonumber \\
  \frac{1}{g'^2}\simeq  \frac{1}{g^2}-\frac{1}{g^2}D(g){\rm
  e}^{-c/g}\cos\theta
\end{eqnarray}
where $c$ is some numerical constant and $D(g)=8\pi^2
mK\sqrt{S_0}$. This equations are qualitatively similar to
$\theta$ and charge renormalization due to instantons in QCD and
$\sigma$ models \cite{KM,P}.

If, on the other hand, $\lambda$ tends to zero, then for the free
motion on the ring $E_0=(1/2m)(\theta/2\pi)^2$ for $\theta<\pi$,
$E_0=(1/2m)(\theta/2\pi-1)^2$ for $\theta>\pi$ and Eqs. (\ref{b}),
(\ref{m}) imply that $m'\rightarrow 0$ while $\theta'\rightarrow
0$, $\theta<\pi$ and $\theta'\rightarrow 2\pi$, $\theta>\pi$.
These results are almost obvious, because at $\lambda=0$ the slow
field ${\bf n}_0$ is no longer coupled to ${\bf n}$.

In the close vicinity of the point $\theta=\pi$ the situation is
more complicated. At $\lambda=0$ the ground state is degenerate,
but the degeneracy is lifted by arbitrarily small external
potential. At small $\lambda$ the energy gap may be expressed as
$\delta E=a\sqrt{\lambda^2+b(\theta-\pi)^2}$, where $a$ and $b$
are some numerical constants, and after expanding in
$(\theta-\pi)$ near the maximum of $E_0(\theta)$ at $\theta=\pi$
we have
\begin{equation}\label{pi}
  E_0(\theta)\simeq {\rm const}
  -\frac{\alpha}{2\lambda}(\theta-\pi)^2,
\end{equation}
where $\alpha=ab$. Hence from Eq. (\ref{m}) $m'\rightarrow
4m^2\pi^2\alpha/\lambda$ at $\lambda\rightarrow 0$ and
\begin{equation}\label{pig}
  1/g'=\sqrt{m'\lambda}\rightarrow 2m\pi\sqrt{\alpha}={\rm
  const},\quad\theta=\pi
\end{equation}
Thus for $\theta=\pi$ the coupling constant $g'$ tends to a fixed
value as $\lambda\rightarrow 0$. This is a kind of quantum
mechanical anomaly (similar to ``rotational anomaly'' of Ref.
\cite{ring}), since strictly at $\lambda=0$ there is no
interaction and $m'$ should be equal to zero. Certainly, for very
small $\lambda$ when $\delta E$ tends to zero near $\theta=\pi$
the adiabatic approximation used here becomes invalid.

Thus the dependence of $m'$ and $\theta'$ on $\lambda$ reproduces
the main features of the famous QHE RG flow diagram. This can be
seen from the Fig.\ref{fig1}, where the evolution of the
renormalized parameters is shown with $\lambda$ decreasing from
top to bottom for different initial values of the vacuum angle
$\theta$. The points in Fig.\ref{fig1} result from numerical
calculation for a simplified model when the term $\lambda\cos\phi$
in Eq.(\ref{H}) is replaced with $\lambda\delta(\phi)$
(qualitative features should not depend on the particular choice
of the potential in Eq. (\ref{H})). Clearly, Fig.(\ref{fig1}) is
similar to the upper half of the QHE RG flow diagram with the
unstable fixed point at $\theta=\pi$ and the ultimate flow of the
renormalized vacuum angle to zero (mod $2\pi$).

\begin{figure}[t]
\includegraphics[width=2in]{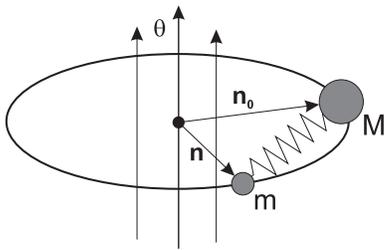}
\caption{Two particles with different masses ($M\gg m$)
interacting via the harmonic potential on the ring with magnetic
flux $\theta$.} \label{fig2}
\end{figure}

The quantum mechanical model discussed here enables, however, a
transparent explanation of why the effective $\theta$ should
vanish as $\lambda\rightarrow 0$. Let us add a kinetic term
$(M/2)\dot{{\bf n}}^2_0$ for the field ${\bf n}_0$ with some large
mass $M$ ($M\gg m$ to ensure the adiabatic approximation) to the
Lagrangian of Eq. (\ref{A}). Then the resulting action with
$w[{\bf n},{\bf n}_0]$ from Eq. (\ref{G}) taken into account
describes two particles with masses $m$ and $M$ interacting via
the harmonic potential, as shown in Fig.\ref{fig2}. Note, that
initially only the light particle interacts with the magnetic flux
$\theta$. One can say that the light particle is charged with,
say, unit charge, while the heavy one is neutral.

Now, if $\lambda$, which determines the interparticle interaction
strength, is high enough, two particles form a tightly bound pair
or an ``atom'', exactly with unit total charge. Mathematically
this means, that the topological term for the field ${\bf n}_0$ is
induced with $\theta'\simeq\theta$ due to the condensation of
charge near the point ${\bf n}_0$. When $\lambda$ decreases, the
bound state gets more loose. When the size of the bound state is
of the order of the ring's radius, rotations of the light particle
are allowed (``instantons'') and its charge is spread along the
ring. So the effective charge of the heavy particle reduces, which
is seen in the formalism as the magnetic flux $\theta$
renormalization.

In summary, we demonstrate how the $\theta$ renormalization may
appear in quantum mechanics of a particle, moving around a thin
ring threaded by a magnetic flux $\theta$. Renormalized $\theta$
is a coefficient in the effective action for the slow variable
${\bf n}_0(\tau)$, which has the meaning of the coarse grained
outcome of the measurement of the particle's position. That is, if
the position is measured with finite accuracy, the observed flux,
equal to the Berry phase associated with the adiabatic rotation of
${\bf n}_0$, will be smaller, than the true one. Formally this
slow variable may be viewed as an additional degree of freedom,
representing a second (heavy) particle on the ring, coupled to the
first one with the harmonic force. Then renormalization of the
flux $\theta$ may be also understood as arising from the change of
the effective charge of the heavy particle when the interaction is
changed.

This example shows, that while the renormalization of the vacuum
angle is definitely a generic property of a system with
instanton-like fluctuations (and the resulting RG flow is not
particularly sensitive to the way the RG transformation is
defined) it does not necessarily  mean that observables are
independent of $\theta$, but is revealed, when the system is being
measured or coupled to some additional slow variable. This
mechanism, leading to small $\theta$ in effective low energy
theory, looks physically different from the direct screening of
$\theta$, as e.g. in the case when the axion field is added, but
it is still not clear whether it has any significance in QCD.

\begin{acknowledgments} The author is grateful to V. Losyakov, A.
Marshakov and especially to A. Morozov for valuable discussions.
The work was supported in part by the RFBR grants No 06-02-17459
and No 07-02-01161.
\end{acknowledgments}


\begin{thebibliography}{18}
\bibitem{Pol} A.M. Polyakov, {\em Gauge Fields and Strings}
(Harwood Academic Publishers, New York, 1987).
 \bibitem{KM}
  V.G. Knizhnik  and A.Yu. Morozov,
 Pis'ma v ZhETF {\bf 39}, 202 (1984). [JETP Lett. {\bf 39}, 240
 (1984)], H. Levine, and S. Libby, Phys. Lett. B {\bf 150}, 182 (1985).
\bibitem{P} H. Levine, S. Libbi, and A.M.M. Pruisken, Nucl. Phys. {\bf B240}
[FS12], 30, 49, 71 (1984), A.M.M. Pruisken,  Nucl. Phys. {\bf
B290}, 61 (1987).
\bibitem{rec} J.I. Latorre and C.A. L\"{u}tken, Phys. Lett. B {\bf 421}, 217 (1998),
 A.M.M. Pruisken, M.A. Baranov, and M. Voropaev, cond-mat/0101003,
 L. Campos Venuti, C. Degli Esposti Boschi, E. Ercolessi, F. Ortolani,
G. Morandi, S. Pasini, and M. Roncaglia, J. Stat. Mech. L02004
(2005), A.M.M. Pruisken, R. Shankar, and N. Surendran, Phys. Rev.
B {\bf 72}, 035329 (2005), A.M.M. Pruisken and I.S. Burmistrov,
Ann. of Phys. (N.Y.) {\bf 316}, 285 (2005).
\bibitem{Sh} For a recent review see G. Gabadadze and M. Shifman,
Int. J. Mod. Phys. A {\bf 17}, 3689 (2002).
\bibitem{Sum} M.A. Shifman, A.I. Vainshtein, and V.I. Zakharov, Nucl. Phys. {\bf B147},
 385 (1979), M.A. Shifman, A.I. Vainshtein, and V.I. Zakharov,
Nucl. Phys. {\bf B166}, 493 (1980).
\bibitem{H} A.M.M. Pruisken, in {\em The Quantum Hall Effect},
eds. R.E. Prange and S. Girvin (Springer, 1990), A.M.M. Pruisken
and I.S. Burmistrov, Ann. Phys. {\bf 322}, 1265 (2007).
\bibitem{B}  S.A. Bulgadaev,
  Pis'ma v ZhETF {\bf 83}, 659 (2006), cond-mat/0605360, I.S.
  Burmistrov and A.M.M. Pruisken, cond-mat/0702400.
\bibitem{A} S.M. Apenko, Phys. Rev. B {\bf 74}, 193311 (2007).
\bibitem{Cep} D.M. Ceperley, Rev. Mod. Phys. {\bf 67}, 279 (1995).
\bibitem{AW} M. Stone, Phys. Rev. D {\bf 33}, 1191 (1986), A.G.
Abanov and P.B. Wigmann, Nucl. Phys. {\bf B570}, 685 (2000).
\bibitem{J} S.M. Apenko, Phys. Lett. A {\bf 142}, 277 (1989),
G. Sch\"{o}n   and A.D. Zaikin  Phys. Rep. {\bf 198}, 237 (1990).
\bibitem{M} R.P. Feynman, Rev. Mod. Phys. {\bf 20}, 367 (1948),
M.B. Mensky, {\em Continuous Quantum Measurements and Path
Integrals} (IOP Publishing, 1993).
\bibitem{N} A.M. Polyakov, Phys. Lett. B {\bf 59}, 79 (1975).
\bibitem{Ber} M. Berry, Proc. Roy. Soc. London, {\bf A392}, 45
(1984).
\bibitem{ring} R. Merlin, Phys. Lett. A {\bf 18}, 421 (1993).
\bibitem{R} R. Rajaraman, {\em Solitons and Instantons}
(North Holland, 1982).
\end{thebibliography}
\end{document}